\documentclass[12pt,a4paper]{article}
\usepackage{fullpage}
\usepackage[centertags]{amsmath}
\allowdisplaybreaks[4]
\numberwithin{equation}{section}
\usepackage{amsbsy}
\usepackage{amsfonts}
\usepackage{amssymb}
\usepackage{url}
\usepackage[dvips]{hyperref}
\usepackage{tocbibind}
\newcommand{\vet}[1]{\ensuremath{\hskip-1pt\vec{\hskip1pt#1}}}
\begin{document}

\begin{flushright}
\textsf{20 February 2004}
\\
\textsf{hep-ph/0402217}
\end{flushright}

\vspace{1cm}

\begin{center}
\large
\textbf{Flavor Neutrinos States}
\normalsize
\\[0.5cm]
\large
Carlo Giunti
\normalsize
\\[0.5cm]
INFN, Sezione di Torino, and Dipartimento di Fisica Teorica,
\\
Universit\`a di Torino,
Via P. Giuria 1, I--10125 Torino, Italy
\\[0.5cm]
\begin{minipage}[t]{0.8\textwidth}
\begin{center}
\textbf{Abstract}
\end{center}
It is shown that a
flavor neutrino state that describes a neutrino produced or detected in
a charged-current weak interaction process
depends on the process under consideration
and is appropriate
for the description of neutrino oscillations
as well as for the calculation of neutrino production or detection rates.
Hence, we have a consistent framework for the description of
neutrino oscillations and interactions.
\end{minipage}
\end{center}

\begin{flushleft}
\end{flushleft}

\tableofcontents

\section{Introduction}
\label{Introduction}

The standard theory of neutrino oscillations
has been derived in the middle 70's
\cite{Eliezer:1976ja,Fritzsch:1976rz,Bilenky:1976yj}
under the assumption that
a neutrino produced or detected in a charged-current weak interaction process
together with a charged lepton with flavor $\alpha$ $=$ $e$, $\mu$ or $\tau$
is described by the flavor state
\begin{equation}
| \nu_{\alpha} \rangle
=
\sum_k U_{\alpha k}^* \, | \nu_{k} \rangle
\,,
\label{1331}
\end{equation}
where $U$ is the unitary mixing matrix of the neutrino fields
and $| \nu_{k} \rangle$ are the Fock states of massive neutrinos
(see the review in Ref.~\cite{Bilenky:1978nj}).
It is then necessary to ask if the flavor state (\ref{1331})
is appropriate also for the description
of the neutrino production and detection rates.
This is a necessary requirement for the
validity of the flavor state (\ref{1331}),
because neutrino production and detection
are essential parts of neutrino oscillation experiments.

In this paper we will show that
the flavor state (\ref{1331})
is appropriate for the description of neutrino production and detection
as well as for the description of neutrino oscillations
in the plane wave approximation
and for experiments which are not sensitive to the
dependence of the interaction probability on the different neutrino masses.
In general
the appropriate flavor state
depends on the process in which the neutrino is produced or detected.
This fact was already noted in Refs.~\cite{Giunti:1992cb,Bilenkii:2001yh,Giunti:2002xg},
where the consequences for neutrino oscillations have been discussed.
Here we will show that
the appropriate flavor state is suitable not only for the description of
neutrino oscillations, but also for the description
of neutrino production or detection.

For definiteness,
we will consider a neutrino produced in the general decay process
\begin{equation}
P_I \to P_F + \ell_{\alpha}^+ + \nu_{\alpha}
\,,
\label{120}
\end{equation}
where $P_I$ and $P_F$ are hadronic or leptonic initial and final particles
and $\ell_{\alpha}^+$ is a positively charged lepton of flavor $\alpha$,
with $\alpha=e,\mu,\tau$.
For example, in the pion decay process
$
\pi^{+} \to \mu^{+} + \nu_{\mu}
$
we have $P_I=\pi^{+}$, $P_F$ is absent and $\alpha=\mu$.
The following considerations can be
easily modified to take into account
a different production process,
as well as a detection process.

In Section~\ref{Plane Wave Approximation}
we consider the flavor neutrino state in the plane wave approximation.
In Subsection~\ref{Plane Wave Approximation: Production Rate}
we show that the flavor neutrino state
can be used in the calculation of the decay rate of the process (\ref{120})
and
in Subsection~\ref{Plane Wave Approximation: Neutrino Oscillations}
we discuss the derivation of neutrino oscillations
in the plane wave approximation.
In Section~\ref{Wave Packet Treatment}
we present the general derivation
in the framework of Quantum Field Theory
of the flavor neutrino state that describes the neutrino produced
in the process (\ref{120})
as a coherent superposition of massive neutrino wave packets,
which is necessary in order to describe
the localization of the production process
and
the related energy-momentum uncertainty which allows
the coherent production of a superposition of different massive neutrinos.
In Subsection~\ref{Wave Packet Treatment: Production Rate}
we show that this flavor neutrino state
leads to the correct decay rate for the process (\ref{120})
and in Subsection~\ref{Wave Packet Treatment: Neutrino Oscillations}
we discuss the implications for neutrino oscillations.
Finally, in Section~\ref{Conclusions}
we present our conclusions.

\section{Plane Wave Approximation}
\label{Plane Wave Approximation}

In neutrino oscillation experiments
the energies and momenta
of the particles which participate to the
neutrino production process
are not measured with a degree of accuracy which would
allow to determine, through energy-momentum conservation,
which massive neutrino
is emitted.
In this case,
a flavor neutrino is a superposition of massive neutrinos.

In the plane wave approach a neutrino with flavor $\alpha$
created in a charged-current weak interaction process
is described by the normalized flavor neutrino state
\cite{Giunti:1992cb,Bilenkii:2001yh}
\begin{equation}
| \nu_{\alpha} \rangle
=
\left( \sum_k |\mathcal{A}_{\alpha k}|^2 \right)^{-1/2}
\sum_k \mathcal{A}_{\alpha k} \, | \nu_{k} \rangle
\,,
\label{1201}
\end{equation}
which is a coherent superposition of massive neutrino states
$ | \nu_{k} \rangle $.
The coefficient
$\mathcal{A}_{\alpha k}$
of the massive neutrino state
is given by the amplitude of production of $\nu_k$,
which,
in general,
depends on the production process.

In the case of the general decay process (\ref{120})
the amplitude $\mathcal{A}_{\alpha k}$ is given by
\begin{equation}
\mathcal{A}_{\alpha k}
=
\langle \nu_k , \ell_{\alpha}^+ , P_F |
\,
\widehat{S}
\,
| P_I \rangle
\,,
\label{1202}
\end{equation}
where $\widehat{S}$ is the $S$-matrix operator.

\subsection{Production Rate}
\label{Plane Wave Approximation: Production Rate}

The amplitude of the general decay process (\ref{120})
is given by
\begin{equation}
\mathcal{A}
=
\langle \nu_{\alpha} , \ell_{\alpha}^+ , P_F |
\,
\widehat{S}
\,
| P_I \rangle
=
\left( \sum_k |\mathcal{A}_{\alpha k}|^2 \right)^{-1/2}
\sum_k \mathcal{A}_{\alpha k}^*
\,
\langle \nu_{k} , \ell_{\alpha}^+ , P_F |
\widehat{S}
| P_I \rangle
=
\left( \sum_k |\mathcal{A}_{\alpha k}|^2 \right)^{1/2}
\,.
\label{1301}
\end{equation}
Therefore,
the decay probability is given by an incoherent
sum of the probabilities of production of different massive neutrinos,
\begin{equation}
|\mathcal{A}|^2
=
\sum_k |\mathcal{A}_{\alpha k}|^2
\,.
\label{1302}
\end{equation}
In other words,
the coherent character of the flavor state (\ref{1201})
is irrelevant for the decay rate.

It is useful to express the $S$-matrix operator as
\begin{equation}
\widehat{S}
=
1
- i \int \mathrm{d}^4x \,
\mathcal{H}_{\mathrm{I}}^{\mathrm{CC}}(x)
\,,
\label{134}
\end{equation}
where we have considered
only the first order perturbative contribution of the
effective low-energy charged-current weak interaction hamiltonian
\begin{align}
\mathcal{H}_{\mathrm{I}}^{\mathrm{CC}}(x)
=
\null & \null
\frac{G_{\mathrm{F}}}{\sqrt{2}}
\sum_{\alpha=e,\mu,\tau}
\overline{\nu_{\alpha}}(x)
\,
\gamma^{\rho}
\left( 1 - \gamma^5 \right)
\ell_{\alpha}(x)
\,
J_{\rho}(x)
+
\text{h.c.}
\nonumber
\\
=
\null & \null
\frac{G_{\mathrm{F}}}{\sqrt{2}}
\sum_{\alpha=e,\mu,\tau}
\sum_{k}
U_{\alpha k}^*
\,
\overline{\nu_{k}}(x)
\,
\gamma^{\rho}
\left( 1 - \gamma^5 \right)
\ell_{\alpha}(x)
\,
J_{\rho}(x)
+
\text{h.c.}
\,.
\label{003}
\end{align}
Here $G_{\mathrm{F}}$ is the Fermi constant
and $J_{\rho}(x)$ is the weak charged current that describes the
transition $P_I \to P_F$.

Taking into account the mixing of the neutrino fields in
the charged-current weak interaction hamiltonian
(\ref{003}),
the amplitude $\mathcal{A}_{\alpha k}$
can be written as
\begin{equation}
\mathcal{A}_{\alpha k}
=
U_{\alpha k}^*
\,
\mathcal{M}_{\alpha k}
\,,
\label{1203}
\end{equation}
where $\mathcal{M}_{\alpha k}$
is the matrix element
\begin{equation}
\mathcal{M}_{\alpha k}
=
- i
\,
\frac{G_{\mathrm{F}}}{\sqrt{2}}
\int \mathrm{d}^4x \,
\langle \nu_k , \ell_{\alpha}^+ , P_F |
\,
\overline{\nu_{k}}(x)
\,
\gamma^{\rho}
\left( 1 - \gamma^5 \right)
\ell_{\alpha}(x)
\,
J_{\rho}(x)
\,
| P_I \rangle
\,.
\label{1204}
\end{equation}

For the decay probability (\ref{1302}) we obtain
\begin{equation}
|\mathcal{A}|^2
=
\sum_k |U_{\alpha k}|^2 \, |\mathcal{M}_{\alpha k}|^2
\,,
\label{13020}
\end{equation}
which is
an incoherent
sum of the probabilities of production of the different massive neutrinos
weighted by $|U_{\alpha k}|^2$,
in agreement with Refs.~\cite{Shrock:1980vy,McKellar:1980cn,Kobzarev:1980nk,Shrock:1981ct,Shrock:1981wq}.

Therefore,
the flavor neutrino state (\ref{1201})
leads to the correct decay rate for the general decay process (\ref{120}).
It is clear that
this proof can be easily generalized to any charged-current
weak interaction process in which flavor neutrinos
are created or destroyed.

If the experiment is not sensitive to the dependence of
$\mathcal{M}_{\alpha k}$
on the different neutrino masses,
it is possible to approximate
\begin{equation}
\mathcal{M}_{\alpha k} \simeq \mathcal{M}_{\alpha}
\,.
\label{13021}
\end{equation}
In this case, since
$ \displaystyle \sum_k |U_{\alpha k}|^2 = 1 $,
we obtain
\begin{equation}
|\mathcal{A}|^2
=
|\mathcal{M}_{\alpha}|^2
\,,
\label{13022}
\end{equation}
which coincides with the standard decay probability for massless neutrinos
if the common scale of neutrino masses is negligible
in comparison with the experimental resolution.

As shown in Ref.~\cite{hep-ph/0302045},
the decay probability (\ref{13022})
can also be obtained starting from the usual flavor state (\ref{1331})
obtained from Eq.~(\ref{1201}) with the approximation (\ref{13021}).
Indeed,
in this case the decay amplitude is given by
\begin{equation}
\mathcal{A}
=
\langle \nu_{\alpha} , \ell_{\alpha}^+ , P_F |
\,
\widehat{S}
\,
| P_I \rangle
=
\sum_k U_{\alpha k} \, \mathcal{A}_{\alpha k}
=
\sum_k |U_{\alpha k}|^2 \, \mathcal{M}_{\alpha}
=
\mathcal{M}_{\alpha}
\,.
\label{1332}
\end{equation}
Let us remark, however,
that a derivation of the decay amplitude starting from the
usual flavor state (\ref{1331})
when the experiment is sensitive to the dependence of $\mathcal{M}_{\alpha k}$
on the different neutrino masses and the approximation (\ref{13021})
is not valid
would lead to a wrong result.

\subsection{Neutrino Oscillations}
\label{Plane Wave Approximation: Neutrino Oscillations}

Let us consider a neutrino oscillation experiments
in which $ \nu_\alpha \to \nu_\beta $
transitions are studied.
Since in this case there are two interaction processes,
one for production (P) and the other for detection (D),
we consider the two flavor neutrino states
\begin{align}
| \nu_{\alpha}^{\mathrm{P}} \rangle
=
\null & \null
\left( \sum_k |\mathcal{A}_{\alpha k}^{\mathrm{P}}|^2 \right)^{-1/2}
\sum_k \mathcal{A}_{\alpha k}^{\mathrm{P}} \, | \nu_{k} \rangle
\,,
\label{1303}
\\
| \nu_{\alpha}^{\mathrm{D}} \rangle
=
\null & \null
\left( \sum_k |\mathcal{A}_{\alpha k}^{\mathrm{D}}|^2 \right)^{-1/2}
\sum_k \mathcal{A}_{\alpha k}^{\mathrm{D}} \, | \nu_{k} \rangle
\,.
\label{1304}
\end{align}
The amplitude of $ \nu_\alpha \to \nu_\beta $
transitions is given by
\begin{equation}
A_{\alpha\beta}(L,T)
=
\langle \nu_{\beta}^{\mathrm{D}} |
e^{ -i \widehat{E} T + i \widehat{P} L }
| \nu_{\alpha}^{\mathrm{P}} \rangle
\,,
\label{13052}
\end{equation}
where
$(L,T)$ is the space-time interval between production and detection
and
$\widehat{E}$ and $\widehat{P}$
are, respectively, the energy and momentum operators.
Since the massive neutrinos have definite masses and kinematical properties,
we obtain
\begin{equation}
A_{\alpha\beta}(L,T)
=
\left( \sum_k |\mathcal{A}_{\alpha k}^{\mathrm{P}}|^2 \right)^{-1/2}
\left( \sum_k |\mathcal{A}_{\beta k}^{\mathrm{D}}|^2 \right)^{-1/2}
\sum_{k}
\mathcal{A}_{\alpha k}^{\mathrm{P}}
\,
\mathcal{A}_{\beta k}^{\mathrm{D}*}
\,
e^{- i E_k T + i p_k L}
\,,
\label{1305}
\end{equation}
with
\begin{equation}
E_k = \sqrt{ p_k^2 + m_k^2 }
\,.
\label{13051}
\end{equation}

In oscillation experiments the neutrino propagation time $T$ is not measured.
In order to express the propagation time $T$ in terms of the known distance
$L$
traveled by the neutrino between production and detection,
we take into account the fact that
neutrinos in oscillation experiments are ultrarelativistic\footnote{
It is known that neutrino masses
are smaller than about one eV
(see Refs.~\cite{Bilenky:2002aw,hep-ph/0310238}).
Since only neutrinos with energy larger than about 100 keV
can be detected (see the discussion in Ref.~\cite{Giunti:2002xg}),
in oscillation experiments neutrinos are always ultrarelativistic.
}.
In this case it is possible to approximate $T \simeq L$,
because
in reality
neutrinos are described by wave packets
\cite{Kayser:1981ye,Giunti:1991ca,hep-ph/9711363,Beuthe:2001rc,Giunti:2002xg,Giunti:2003ax},
which are localized on the production process
at the production time and
propagate between the production and detection processes with a group velocity
close to the velocity of light.

The physical reason why the approximation $T \simeq L$ is correct
can be understood by noting that,
if the massive neutrinos are ultrarelativistic
and contribute coherently to the detection process,
their wave packets
overlap with the detection process
for an interval of time $[ t - \Delta t \,,\, t + \Delta t ]$,
with
\begin{equation}
t
=
\frac{L}{\overline{v}}
\simeq
L \left( 1 + \frac{\overline{m^2}}{2E^2} \right)
\,,
\qquad
\Delta t \sim \sigma_x
\,,
\label{505}
\end{equation}
where $\overline{v}$
is the average group velocity,
$\overline{m^2}$ is the average of the squared neutrino masses,
$\sigma_x$
is given by the spatial uncertainties of the production and detection processes
summed in quadrature
\cite{hep-ph/9711363}
(the spatial uncertainty of the production process
determines the size of the massive neutrino wave packets).
The correction $ L \overline{m^2} / 2E^2 $ to $t=L$
in Eq.~(\ref{505})
can be neglected,
because it gives corrections to the oscillation phases
which are of higher order in the very small ratios
$ m_k^2 / E^2 $.
The corrections due to
$\Delta t \sim \sigma_x$
are also negligible,
because in all realistic experiments
$ \sigma_x $
is much smaller than the oscillation length
$L^{\mathrm{osc}}_{kj} = 4 \pi E / \Delta{m}^2_{kj}$,
otherwise oscillations could not be observed
\cite{Kayser:1981ye,Giunti:1991ca,Beuthe:2001rc,Giunti:2003ax}.
One can summarize these arguments by saying that
the approximation $T \simeq L$ is correct
because
the phase of the oscillations
is practically constant over the interval of time in which the
massive neutrino wave packets overlap with the detection process.

Using the approximation
$T \simeq L$
the phase in Eq.~(\ref{1305}) becomes
\begin{equation}
- E_k T + p_k L
\simeq
- \left( E_k - p_k \right) L
=
- \frac{ E_k^2 - p_k^2 }{ E_k + p_k } \, L
=
- \frac{ m_k^2 }{ E_k + p_k } \, L
\simeq
- \frac{ m_k^2 }{ 2 E } \, L
\,,
\label{1005}
\end{equation}
where $E$ is the neutrino energy neglecting mass contributions.
It is important to notice that Eq.~(\ref{1005})
shows that the phases of massive neutrinos relevant for the oscillations
are independent from the particular values of the energies and momenta
of different massive neutrinos
\cite{Winter:1981kj,Giunti:1991ca,Giunti:2000kw,Giunti:2001kj,Giunti:2003ax},
as long as the relativistic dispersion relation (\ref{13051})
is satisfied.

The probability of
$ \nu_\alpha \to \nu_\beta $
transitions in space is given by
\begin{equation}
P_{\alpha\beta}(L)
\simeq
\left( \sum_k |\mathcal{A}_{\alpha k}^{\mathrm{P}}|^2 \right)
\left( \sum_k |\mathcal{A}_{\beta k}^{\mathrm{D}}|^2 \right)
\sum_{k,j}
\mathcal{A}_{\alpha k}^{\mathrm{P}}
\,
\mathcal{A}_{\beta k}^{\mathrm{D}*}
\,
\mathcal{A}_{\alpha j}^{\mathrm{P}*}
\,
\mathcal{A}_{\beta j}^{\mathrm{D}}
\,
\exp\left( - i \frac{\Delta{m}^2_{kj}L}{2E} \right)
\,,
\label{1306}
\end{equation}
with
$\Delta{m}^2_{kj} \equiv m_k^2 - m_j^2$.

Oscillations can be observed at a macroscopic distance
$L$
only if the difference between two neutrino masses
is much smaller than the neutrino energy $E$.
For example,
the oscillation length
$
L^{\mathrm{osc}}_{kj}
=
4 \pi E / \Delta{m}^2_{kj}
$
is larger than about 1 m
if
$\Delta{m}^2_{kj} \lesssim 2.5 \, \mathrm{eV}^2$
for
$ E \simeq 1 \, \mathrm{MeV} $.
In this case
the difference of neutrino masses can be neglected
in the production and detection amplitudes\footnote{
We implicitly assume also that energy-momentum conservation allows
the coherent production and detection of massive neutrinos.
See the discussion after Eq.(\ref{202}).
}:
\begin{equation}
\mathcal{A}_{\alpha k}^{\mathrm{P}}
\simeq
U_{\alpha k}^* \, \mathcal{M}_{\alpha}^{\mathrm{P}}
\,,
\qquad
\mathcal{A}_{\beta k}^{\mathrm{D}}
\simeq
U_{\beta k}^* \, \mathcal{M}_{\beta}^{\mathrm{D}}
\,,
\label{1307}
\end{equation}
where
$\mathcal{M}_{\alpha}^{\mathrm{P}}$
is the matrix element (\ref{1204})
in which the difference of neutrino masses has been neglected,
and $\mathcal{M}_{\beta}^{\mathrm{D}}$
is a similar matrix element for the detection process.
Taking into account these approximations,
the transition probability (\ref{1306})
reduces to the standard one
\cite{Eliezer:1976ja,Fritzsch:1976rz,Bilenky:1976yj,Bilenky:1978nj},
\begin{equation}
P_{\alpha\beta}(L)
\simeq
\sum_{k,j}
U_{\alpha k}^*
\,
U_{\beta k}
\,
U_{\alpha j}
\,
U_{\beta j}^*
\,
\exp\left( - i \frac{\Delta{m}^2_{kj}L}{2E} \right)
\,,
\label{13071}
\end{equation}
which can be obtained starting from the standard
production and detection flavor states
\begin{equation}
| \nu_{\alpha}^{\mathrm{P}} \rangle
=
\sum_k U_{\alpha k}^* \, | \nu_{k} \rangle
\,,
\qquad
| \nu_{\beta}^{\mathrm{D}} \rangle
=
\sum_k U_{\beta k}^* \, | \nu_{k} \rangle
\,,
\label{1308}
\end{equation}
obtained from Eqs.~(\ref{1303}) and (\ref{1304})
through the approximations (\ref{1307}).
Therefore,
the standard flavor states in Eq.~(\ref{1331}) and (\ref{1308})
are appropriate to describe neutrino oscillations
in the plane wave approximation,
taking into account that in all neutrino oscillation experiments
the
dependence of the production and detection probabilities on the different neutrino masses
is negligible.

\section{Wave Packet Treatment}
\label{Wave Packet Treatment}

In Quantum Field Theory the asymptotic final state resulting from the decay
of the initial particle $P_I$ in Eq.~(\ref{120})
is given by
\begin{equation}
| f \rangle
=
\widehat{S} \, | P_I \rangle
\,.
\label{122}
\end{equation}
This final state
is an entangled state in which the final particles do not have
individual separate properties.
However,
in practice the decay always occur in a medium
where
$P_F$
and
$\ell_{\alpha}^+$
interact very quickly,
reducing the final state to a disentangled state
$ | \nu_{\alpha} , \ell_{\alpha}^+ , P_F \rangle $
in which each particle has individual properties.
Hence, the final flavor neutrino state is given by
\cite{Giunti:2002xg}
\begin{equation}
| \nu_{\alpha} \rangle
\propto
\langle \ell_{\alpha}^+ , P_F | f \rangle
=
\langle \ell_{\alpha}^+ , P_F |
\,
\widehat{S}
\,
| P_I \rangle
\,.
\label{124}
\end{equation}
The proportionality sign in Eq.~(\ref{124})
is necessary in order to take into account the normalization of the flavor neutrino state.

Inserting a completeness on the left of the right-hand side of Eq.~(\ref{124}),
we obtain
\begin{equation}
| \nu_{\alpha} \rangle
\propto
\sum_k \int \mathrm{d}^3p \sum_h
| \nu_k(\vet{p},h) \rangle
\langle \nu_k(\vet{p},h) , \ell_{\alpha}^+ , P_F |
\,
\widehat{S}
\,
| P_I \rangle
\,,
\label{131}
\end{equation}
where $\vet{p}$ is the neutrino momentum and $h$ is the neutrino helicity.
The normalized flavor neutrino state can be written as
\begin{equation}
| \nu_{\alpha} \rangle
=
\left( \sum_k \int \mathrm{d}^3p \sum_h |\mathcal{A}_{\alpha k}(\vet{p},h)|^2 \right)^{-1/2}
\sum_k \int \mathrm{d}^3p \sum_h \mathcal{A}_{\alpha k}(\vet{p},h)
\,
| \nu_{k}(\vet{p},h) \rangle
\,,
\label{201}
\end{equation}
which is a coherent superposition of massive neutrino wave packets.
The coefficient
$\mathcal{A}_{\alpha k}(\vet{p},h)$
is given by the amplitude of production of $\nu_k(\vet{p},h)$:
\begin{equation}
\mathcal{A}_{\alpha k}(\vet{p},h)
=
\langle \nu_k(\vet{p},h) , \ell_{\alpha}^+ , P_F |
\,
\widehat{S}
\,
| P_I \rangle
\,.
\label{202}
\end{equation}

It is important to notice that
if all the particles
$P_I$,
$P_F$,
$\ell_{\alpha}^+$
are described by plane waves,
the production process is not localized and
energy-momentum conservation forbids the coherent production of
different massive neutrinos.
Therefore,
in order to have neutrino oscillations
the particles
$P_I$,
$P_F$,
$\ell_{\alpha}^+$
must be described by localized wave packets
with sufficient energy-momentum uncertainty.
If the energy-momentum uncertainty
is so small that different massive neutrinos
cannot be produced coherently,
the state (\ref{201})
becomes effectively an incoherent mixture of massive neutrino wave packets,
because the different energy-momentum conservations contained in the amplitudes
(\ref{202})
cannot be satisfied simultaneously.

In the discussion of the
plane wave approximation in
Section~\ref{Plane Wave Approximation}
we swept this problem under the carpet.
However,
there are no implications for the neutrino production and detection rates
discussed in Subsection~\ref{Plane Wave Approximation: Production Rate}
which,
as we have seen in Eq.~(\ref{1302}),
are given
by an incoherent
sum of the probabilities of production or detection of the different massive neutrinos.
On the other hand,
the oscillation probability (\ref{1306})
has, strictly speaking,
no physical meaning,
because the exact energy-momentum conservation delta-functions
contained in the amplitudes (\ref{1202})
for different massive neutrinos
are mutually exclusive.
However,
these delta-functions have been factorized out of the
standard oscillation probability (\ref{13071}),
which acquires a physical meaning as the approximation of
the oscillation probability in the limit of
negligible wave packet effects,
as we will see in Subsection~\ref{Wave Packet Treatment: Neutrino Oscillations}.

\subsection{Production Rate}
\label{Wave Packet Treatment: Production Rate}

The amplitude of the general decay process (\ref{120})
is given by
\begin{align}
\mathcal{A}
=
\null & \null
\langle \nu_{\alpha} , \ell_{\alpha}^+ , P_F |
\,
\widehat{S}
\,
| P_I \rangle
\nonumber
\\
=
\null & \null
\left( \sum_k \int \mathrm{d}^3p \sum_h |\mathcal{A}_{\alpha k}(\vet{p},h)|^2 \right)^{-1/2}
\sum_k \int \mathrm{d}^3p \sum_h \mathcal{A}_{\alpha k}^*(\vet{p},h)
\,
\langle \nu_k(\vet{p},h) , \ell_{\alpha}^+ , P_F |
\,
\widehat{S}
\,
| P_I \rangle
\nonumber
\\
=
\null & \null
\left( \sum_k \int \mathrm{d}^3p \sum_h |\mathcal{A}_{\alpha k}(\vet{p},h)|^2 \right)^{1/2}
\,.
\label{301}
\end{align}
Hence,
the decay probability is given by
\begin{equation}
|\mathcal{A}|^2
=
\sum_k \int \mathrm{d}^3p \sum_h |\mathcal{A}_{\alpha k}(\vet{p},h)|^2
\,,
\label{132}
\end{equation}
which is an incoherent sum
of the probabilities of production of the different massive neutrinos.

Using the first order perturbative expansion of the $S$-matrix operator in Eq.~(\ref{134}),
the amplitudes
$\mathcal{A}_{\alpha k}(\vet{p},h)$
can be written as
\begin{equation}
\mathcal{A}_{\alpha k}(\vet{p},h)
=
U_{\alpha k}^*
\,
\mathcal{M}_{\alpha k}(\vet{p},h)
\,,
\label{133}
\end{equation}
where
$ \mathcal{M}_{\alpha k}(\vet{p},h) $
are the matrix elements
\begin{equation}
\mathcal{M}_{\alpha k}(\vet{p},h)
=
- i
\,
\frac{G_{\mathrm{F}}}{\sqrt{2}}
\int \mathrm{d}^4x \,
\langle \nu_k(\vet{p},h) , \ell_{\alpha}^+ , P_F |
\,
\overline{\nu_{k}}(x)
\,
\gamma^{\rho}
\left( 1 - \gamma^5 \right)
\ell_{\alpha}(x)
\,
J_{\rho}(x)
\,
| P_I \rangle
\,.
\label{204}
\end{equation}

Using the expression (\ref{133}) in Eq.~(\ref{132})
it becomes clear that,
in agreement with Refs.~\cite{Shrock:1980vy,McKellar:1980cn,Kobzarev:1980nk,Shrock:1981ct,Shrock:1981wq},
the decay probability is given by an incoherent
sum of the probabilities
of production of the different massive neutrinos
weighted by $|U_{\alpha k}|^2$,
\begin{equation}
|\mathcal{A}|^2
=
\sum_k |U_{\alpha k}|^2 \int \mathrm{d}^3p \sum_h |\mathcal{M}_{\alpha k}(\vet{p},h)|^2
\,.
\label{302}
\end{equation}

Therefore, also in a quantum field theoretical wave packet treatment,
a description of the flavor neutrino created in the process (\ref{120})
with the process-dependent coherent state (\ref{201})
leads to the correct decay rate,
for which the coherent character of the superposition
on massive neutrinos is irrelevant.
This proof can be easily generalized to any charged-current
weak interaction process in which flavor neutrinos
are created or destroyed.

If the experiment is not sensitive to the dependence of
$\mathcal{M}_{\alpha k}(\vet{p},h)$
on the neutrino masses,
it is possible to approximate
\begin{equation}
\mathcal{M}_{\alpha k}(\vet{p},h) \simeq \mathcal{M}_{\alpha}(\vet{p},h)
\,.
\label{3021}
\end{equation}
In this case
we obtain
\begin{equation}
|\mathcal{A}|^2
=
\int \mathrm{d}^3p \sum_h |\mathcal{M}_{\alpha}(\vet{p},h)|^2
\,,
\label{3022}
\end{equation}
which coincides with the decay amplitude for massless neutrinos
in the wave packet treatment
if the common scale of neutrino masses is negligible
in comparison with the experimental resolution.

\subsection{Neutrino Oscillations}
\label{Wave Packet Treatment: Neutrino Oscillations}

Let us consider the two 
production (P) and detection (D)
neutrino states
\begin{align}
| \nu_{\alpha}^{\mathrm{P}} \rangle
=
N_{\alpha}^{\mathrm{P}}
\sum_k \int \mathrm{d}^3p \sum_h \mathcal{A}_{\alpha k}^{\mathrm{P}}(\vet{p},h)
\,
| \nu_{k}(\vet{p},h) \rangle
\,,
\label{303}
\\
| \nu_{\alpha}^{\mathrm{D}} \rangle
=
N_{\alpha}^{\mathrm{D}}
\sum_k \int \mathrm{d}^3p \sum_h \mathcal{A}_{\alpha k}^{\mathrm{D}}(\vet{p},h)
\,
| \nu_{k}(\vet{p},h) \rangle
\,,
\label{304}
\end{align}
with the normalization factors
\begin{equation}
N_{\alpha}^{\mathrm{I}}
=
\left( \sum_k \int \mathrm{d}^3p \sum_h |\mathcal{A}_{\alpha k}^{\mathrm{I}}(\vet{p},h)|^2 \right)^{-1/2}
\,,
\label{300}
\end{equation}
for $\mathrm{I}=\mathrm{P},\mathrm{D}$.
The amplitude of $ \nu_\alpha \to \nu_\beta $
transitions is given by
\begin{equation}
A_{\alpha\beta}(\vet{L},T)
=
\langle \nu_{\beta}^{\mathrm{D}} |
e^{ -i \widehat{E} T + i \widehat{\vet{P}} \cdot \vet{L} }
| \nu_{\alpha}^{\mathrm{P}} \rangle
\,,
\label{305}
\end{equation}
where
$(\vet{L},T)$ is the space-time interval between production and detection
and
$\widehat{E}$ and $\widehat{\vet{P}}$
are, respectively, the energy and momentum operators.
Using the flavor states (\ref{303}) and (\ref{304}) we obtain
\begin{equation}
A_{\alpha\beta}(\vet{L},T)
=
N_{\alpha}^{\mathrm{P}}
\,
N_{\beta}^{\mathrm{D}}
\sum_k \int \mathrm{d}^3p \sum_h
\mathcal{A}_{\alpha k}^{\mathrm{P}}(\vet{p},h)
\,
\mathcal{A}_{\beta k}^{\mathrm{D}*}(\vet{p},h)
\,
e^{- i E_k(\vet{p}) T + i \vet{p} \cdot \vet{L}}
\,,
\label{306}
\end{equation}
with
\begin{equation}
E_k(\vet{p}) = \sqrt{ \vet{p} + m_k^2 }
\,.
\label{307}
\end{equation}
The derivation of the neutrino oscillation probability
from the explicit values of the production and detection amplitudes
in Eq.~(\ref{307})
has been discussed in Ref.~\cite{Giunti:2002xg}.
Since it is rather complicated,
here we consider the approximation
\begin{equation}
\mathcal{A}_{\alpha k}^{\mathrm{P}}(\vet{p},h)
\,
\mathcal{A}_{\beta k}^{\mathrm{D}*}(\vet{p},h)
\simeq
U_{\alpha k}^*
\,
U_{\beta k}
\,
\widetilde{\mathcal{M}}_{\alpha k}^{\mathrm{P}}(\vet{p},h)
\,
\widetilde{\mathcal{M}}_{\beta k}^{\mathrm{D}*}(\vet{p},h)
\,
\exp\left[
- \frac{(\vet{p}-\vet{p}_k)^2}{4\sigma_p^2}
\right]
\,,
\label{308}
\end{equation}
where $\vet{p}_k$
is the average momentum of the $k^{\mathrm{th}}$ massive neutrino component
and
the exponential takes into account
energy-momentum conservation
within the momentum uncertainty
$\sigma_p$
determined by the widths of the momentum distributions
of the wave packets of the particles
participation to the production and detection processes
(see Ref.~\cite{Giunti:2002xg}).
We choose a gaussian form for the momentum distribution in order to
be able to perform the integral over $\mathrm{d}^3p$
analitically.
This approximation is almost equivalent to the saddle-point approximation
performed in Ref.~\cite{Giunti:2002xg}
(it is equivalent if $\omega=1$, with $\omega$ defined in Ref.~\cite{Giunti:2002xg}).

Furthermore,
in order to simplify the derivation of the
oscillation probability for realistic experimental setups as much as possible,
we consider ultrarelativistic neutrinos and we
adopt the following assumptions which in practice are always verified:
a) $\sigma_p \ll p_k$, with $p_k=|\vet{p}_k|$;
b) $
\widetilde{\mathcal{M}}_{\alpha k}^{\mathrm{P}}(\vet{p},h)
\,
\widetilde{\mathcal{M}}_{\beta k}^{\mathrm{D}*}(\vet{p},h)
$
is a smooth function of $\vet{p}$;
c) the experiment is not sensitive to the dependence of
$
\widetilde{\mathcal{M}}_{\alpha k}^{\mathrm{P}}(\vet{p},h)
\,
\widetilde{\mathcal{M}}_{\beta k}^{\mathrm{D}*}(\vet{p},h)
$
on the different neutrino masses.
Under these assumptions we can approximate
\begin{equation}
\widetilde{\mathcal{M}}_{\alpha k}^{\mathrm{P}}(\vet{p},h)
\,
\widetilde{\mathcal{M}}_{\beta k}^{\mathrm{D}*}(\vet{p},h)
\simeq
\widetilde{\mathcal{M}}_{\alpha}^{\mathrm{P}}(\vet{p}_{\nu},h)
\,
\widetilde{\mathcal{M}}_{\beta}^{\mathrm{D}*}(\vet{p}_{\nu},h)
\,,
\label{309}
\end{equation}
where $\vet{p}_{\nu}$ is the neutrino momentum neglecting mass effects.
The energy $E_k(\vet{p})$
can be approximated by
\begin{equation}
E_k(\vet{p})
\simeq
E_k + \vet{v}_k \left( \vet{p} - \vet{p}_k \right)
\,,
\label{310}
\end{equation}
with $E_k$ given by Eq.~(\ref{13051})
and
\begin{equation}
\vet{v}_k
=
\left.
\frac{\partial E_k(\vet{p})}{\partial \vet{p}}
\right|_{\vet{p} = \vet{p}_k}
=
\frac{\vet{p}_k}{E_k}
\simeq
1 - \frac{m_k^2}{2E^2}
\,,
\label{311}
\end{equation}
where
$ E = |\vet{p}_{\nu}| $
is the neutrino energy neglecting mass effects.
Since,
as shown in Ref.~\cite{Giunti:2002xg},
in the case of ultrarelativistic neutrinos
all the momenta $\vet{p}_k$
are aligned in the direction of $L$,
with these approximations,
the transition amplitude (\ref{306})
reduces to
\begin{align}
A_{\alpha\beta}(L,T)
\simeq
\null & \null
N_{\alpha}^{\mathrm{P}}
\,
N_{\beta}^{\mathrm{D}}
\sum_h
\widetilde{\mathcal{M}}_{\alpha}^{\mathrm{P}}(\vet{p}_{\nu},h)
\,
\widetilde{\mathcal{M}}_{\beta}^{\mathrm{D}*}(\vet{p}_{\nu},h)
\sum_k
U_{\alpha k}^*
\,
U_{\beta k}
\,
e^{ - i E_k T + i p_k L }
\nonumber
\\
\null & \null
\hspace{2cm}
\times
\int \mathrm{d}^3p
\exp\left[
- \frac{(\vet{p}-\vet{p}_k)^2}{4\sigma_p^2}
+ i \left(\vet{p}-\vet{p}_k\right) (\vet{L}-\vet{v}_kT)
\right]
\nonumber
\\
\propto
\null & \null
\sum_k
U_{\alpha k}^*
\,
U_{\beta k}
\,
\exp\left[
- i E_k T + i p_k L
- \frac{(\vet{L}-\vet{v}_kT)^2}{4\sigma_x^2}
\right]
\,,
\label{312}
\end{align}
with the spatial uncertainty $\sigma_x$
related to the momentum uncertainty $\sigma_p$
by the minimal Heisenberg relation
\begin{equation}
\sigma_x \, \sigma_p = \frac{1}{2}
\,.
\label{361}
\end{equation}

In order to obtain the oscillation probability as a function of the
known distance $L$
traveled by the neutrino between production and detection,
the probability
$P_{\alpha\beta}(L,T)=|A_{\alpha\beta}(L,T)|^2$
must be integrated over the unknown time $T$
\cite{Giunti:1991ca}.
Since the integral over $T$
is gaussian, we easily obtain
\begin{align}
P_{\alpha\beta}(L)
\simeq
\null & \null
\sum_{k,j}
U_{\alpha k}^*
\,
U_{\beta k}
\,
U_{\alpha j}
\,
U_{\beta j}^*
\,
\exp\left\{
- i
\left[
\left( E_k - E_j \right)
- \left( p_k - p_j \right)
\right]
L
\right\}
\nonumber
\\
\null & \null
\hspace{2cm}
\times
\exp\left\{
- \left( \frac{\Delta{m}^2_{kj} L}{4 \sqrt{2} E^2 \sigma_x} \right)^2
- \left( \frac{E_k-E_j}{2 \sqrt{2} \sigma_p} \right)^2
\right\}
\,.
\label{313}
\end{align}
Using the same method as in Eq.~(\ref{1005}),
the phase
$
\left[
\left( E_k - E_j \right)
- \left( p_k - p_j \right)
\right]
L
$
becomes the standard oscillation phase
$ \Delta{m}^2_{kj} L / 2 E $
in Eq.~(\ref{13071}).
Since
for ultrarelativistic neutrinos
the energies $E_k$ can be written as
\cite{Giunti:1991ca,Giunti:2000kw,Giunti:2003ax}
\begin{equation}
E_k \simeq E + \xi \, \frac{m_k^2}{2E}
\,,
\label{353}
\end{equation}
where
$\xi$ is a number, usually of order one
(see Ref.~\cite{Giunti:2002xg}),
which depends on the details of the
production and detection processes,
the oscillation probability can be written as
\begin{equation}
P_{\alpha\beta}(L)
=
\sum_{k,j}
U_{\alpha k}^* U_{\alpha j} U_{\beta k} U_{\beta j}^*
\exp\left[
- 2 \pi i \frac{ L }{ L^{\mathrm{osc}}_{kj} }
-
\left(
\frac{ L }{ L^{\mathrm{coh}}_{kj} }
\right)^2
-
2 \pi^2 \xi^2
\left(
\frac{ \sigma_x }{ L^{\mathrm{osc}}_{kj} }
\right)^2
\right]
\,,
\label{017}
\end{equation}
with
the standard oscillation lengths
\begin{equation}
L^{\mathrm{osc}}_{kj}
=
\frac{ 4 \pi E }{ \Delta{m}^2_{kj} }
\,,
\label{0181}
\end{equation}
and the coherence lengths
\cite{Nussinov:1976uw,Kiers:1996zj}
\begin{equation}
L^{\mathrm{coh}}_{kj}
=
\frac{ 4 \sqrt{2} E^2 }{ |\Delta{m}^2_{kj}| }
\sigma_x
\,.
\label{0182}
\end{equation}
As promised at the end of the introduction of Section~\ref{Wave Packet Treatment},
in the limit of
negligible wave packet effects,
\emph{i.e.} for
$L \ll L^{\mathrm{coh}}_{kj}$
and
$\sigma_x \ll L^{\mathrm{osc}}_{kj}$,
the oscillation probability in the wave packet approach reduces to the standard one
in Eq.~(\ref{13071}),
obtained in the plane wave approximation.

The physical meaning of the coherence and localization terms
which appear in Eq.~(\ref{017})
in addition to the standard oscillation phase
have been already discussed at length in
Refs.~\cite{Giunti:1991ca,hep-ph/9711363,Beuthe:2001rc,Giunti:2002xg,Giunti:2003ax}
(see also Refs.~\cite{hep-ph/9305276,hep-ph/9709494,hep-ph/9909332,Beuthe:2002ej}).

In particular,
the localization term
$ \displaystyle
\exp[- 2 \pi^2 \xi^2 ( \sigma_{x} / L^{\mathrm{osc}}_{kj} )^2]
$
suppresses the oscillations due to
$\Delta{m}^2_{kj}$
if
$ \sigma_{x} \gtrsim L^{\mathrm{osc}}_{kj} $.
This means that in order to measure the interference
of the massive neutrino components
$\nu_k$ and $\nu_j$
the production and detection processes must be localized
in space-time regions much smaller than the
oscillation length $L^{\mathrm{osc}}_{kj}$.
In practice this requirement is satisfied in all
neutrino oscillation experiments.

The localization term allows to distinguish
neutrino oscillation experiments
from experiments on the measurement of neutrino masses.
As first shown in Ref.~\cite{Kayser:1981ye},
neutrino oscillations are suppressed
in experiments which are
able to measure,
through energy-momentum conservation,
the mass of the neutrino.
Indeed,
from the energy-momentum dispersion relation (\ref{13051})
the uncertainty of the mass determination is
\begin{equation}
\delta{m_k}^2
=
\sqrt{ \left( 2 E_k \delta{E_k} \right)^2 + \left( 2 p_k \delta{p_k} \right)^2 }
\simeq
2 \sqrt{2} E \sigma_p
\,,
\label{022}
\end{equation}
where the approximation holds for realistic ultrarelativistic neutrinos.
If
$
\delta{m_k}^2
<
|\Delta{m}^2_{kj}|
$,
the mass of $\nu_k$ is measured
with an accuracy better than the difference
$\Delta{m}^2_{kj}$.
In this case
the neutrino $\nu_j$ is not produced or detected
and the interference of
$\nu_k$ and $\nu_j$
which would generate oscillations does not occur.
The localization term
in the oscillation probability
(\ref{017})
automatically 
suppresses
the interference of
$\nu_k$ and $\nu_j$,
because it can be written as
\begin{equation}
\exp\left[
-
2 \pi^2 \xi^2
\left(
\frac{ \sigma_x }{ L^{\mathrm{osc}}_{kj} }
\right)^2
\right]
=
\exp\left[
- \xi^2
\left(
\frac{ \Delta{m}^2_{kj} }{ 4 \sqrt{2} E \sigma_p }
\right)^2
\right]
\simeq
\exp\left[
- \frac{\xi^2}{4}
\left(
\frac{ \Delta{m}^2_{kj} }{ \delta{m_k}^2 }
\right)^2
\right]
\,.
\label{362}
\end{equation}
It is important to notice,
however,
that the validity of this interpretation of the localization term
hinges on the validity of the flavor states (\ref{303}) and (\ref{304})
not only for the description of neutrino oscillations
but also for the description of
neutrino production and detection,
which we have proved in the previous Section.
The suppression of oscillations due to the localization term
reflects the effective loss of coherence of the flavor states
discussed after Eq.(\ref{202}).

\section{Conclusions}
\label{Conclusions}

We have presented a consistent framework for the description of
neutrino oscillations and interactions.

We have shown that the
flavor neutrino state that describes a neutrino produced or detected in
a charged-current weak interaction process
depends on the process under consideration
and is appropriate for the description of neutrino oscillations
as well as for the calculation of neutrino production and detection rates.
We have proved these facts both in the
plane wave approximation
(Section~\ref{Plane Wave Approximation})
and
in the quantum field theoretical wave packet treatment
(Section~\ref{Wave Packet Treatment}).

In the plane wave approximation
the flavor neutrino state can be approximated with
the standard expression in Eq.~(\ref{1331})
only for experiments
which are not sensitive to the
dependence of the interaction probability on the different neutrino masses.
This occurs in all neutrino oscillation experiments.

In the quantum field theoretical wave packet treatment
the flavor neutrino state
takes into account the localization of the neutrino production or detection
process
and the related energy-momentum uncertainty.
The oscillation probability depends on the standard oscillation phase
plus additional coherence and localization terms due to the
wave packet treatment.
The localization term
suppresses the oscillations
if the energy-momentum uncertainty is so small that only one massive neutrino
is produced.
This occurs in experiments on the measurement of masses,
whose neutrino production rate
follow consistently from the same flavor neutrino states
employed for the calculation of the oscillation probability.

Finally,
we would like to remark that
the validity of the process-dependent flavor neutrino states
is consistent with the proof presented in Ref.~\cite{hep-ph/0312256}
that flavor neutrino Fock spaces
\cite{Blasone:1995zc,Blasone:1998hf,Fujii:1998xa}
are clever mathematical constructs
without physical relevance.

\section*{Acknowledgments}

I would like to thank
Kanji Fujii,
Kenzo Ishikawa,
Peter Minkowski
and
Takashi Shimomura
for
very stimulating discussions.


\begin{thebibliography}{10}

\bibitem{Eliezer:1976ja}
S. Eliezer and A.R. Swift,
\newblock Nucl. Phys. B105 (1976) 45.

\bibitem{Fritzsch:1976rz}
H. Fritzsch and P. Minkowski,
\newblock Phys. Lett. B62 (1976) 72.

\bibitem{Bilenky:1976yj}
S.M. Bilenky and B. Pontecorvo,
\newblock Nuovo Cim. Lett. 17 (1976) 569.

\bibitem{Bilenky:1978nj}
S.M. Bilenky and B. Pontecorvo,
\newblock Phys. Rept. 41 (1978) 225.

\bibitem{Giunti:1992cb}
C. Giunti, C.W. Kim and U.W. Lee,
\newblock Phys. Rev. D45 (1992) 2414.

\bibitem{Bilenkii:2001yh}
S.M. Bilenky and C. Giunti,
\newblock Int. J. Mod. Phys. A16 (2001) 3931,
  \href{http://arxiv.org/abs/hep-ph/0102320}{\url{hep-ph/0102320}}.

\bibitem{Giunti:2002xg}
C. Giunti,
\newblock JHEP 11 (2002) 017,
  \href{http://arxiv.org/abs/hep-ph/0205014}{\url{hep-ph/0205014}}.

\bibitem{Shrock:1980vy}
R.E. Shrock,
\newblock Phys. Lett. B96 (1980) 159.

\bibitem{McKellar:1980cn}
B.H.J. McKellar,
\newblock Phys. Lett. B97 (1980) 93.

\bibitem{Kobzarev:1980nk}
I.Y. Kobzarev et~al.,
\newblock Sov. J. Nucl. Phys. 32 (1980) 823.

\bibitem{Shrock:1981ct}
R.E. Shrock,
\newblock Phys. Rev. D24 (1981) 1232.

\bibitem{Shrock:1981wq}
R.E. Shrock,
\newblock Phys. Rev. D24 (1981) 1275.

\bibitem{hep-ph/0302045}
C. Giunti,
\newblock (2003),
  \href{http://arxiv.org/abs/hep-ph/0302045}{\url{hep-ph/0302045}}.

\bibitem{Bilenky:2002aw}
S.M. Bilenky et~al.,
\newblock Phys. Rept. 379 (2003) 69,
  \href{http://arxiv.org/abs/hep-ph/0211462}{\url{hep-ph/0211462}}.

\bibitem{hep-ph/0310238}
C. Giunti and M. Laveder,
\newblock (2003),
  \href{http://arxiv.org/abs/hep-ph/0310238}{\url{hep-ph/0310238}}.

\bibitem{Kayser:1981ye}
B. Kayser,
\newblock Phys. Rev. D24 (1981) 110.

\bibitem{Giunti:1991ca}
C. Giunti, C.W. Kim and U.W. Lee,
\newblock Phys. Rev. D44 (1991) 3635.

\bibitem{hep-ph/9711363}
C. Giunti and C.W. Kim,
\newblock Phys. Rev. D58 (1998) 017301,
  \href{http://arxiv.org/abs/hep-ph/9711363}{\url{hep-ph/9711363}}.

\bibitem{Beuthe:2001rc}
M. Beuthe,
\newblock Phys. Rept. 375 (2003) 105,
  \href{http://arxiv.org/abs/hep-ph/0109119}{\url{hep-ph/0109119}}.

\bibitem{Giunti:2003ax}
C. Giunti,
\newblock Found. Phys. Lett. 17 (2004) 103,
  \href{http://arxiv.org/abs/hep-ph/0302026}{\url{hep-ph/0302026}}.

\bibitem{Winter:1981kj}
R.G. Winter,
\newblock Lett. Nuovo Cim. 30 (1981) 101.

\bibitem{Giunti:2000kw}
C. Giunti and C.W. Kim,
\newblock Found. Phys. Lett. 14 (2001) 213,
  \href{http://arxiv.org/abs/hep-ph/0011074}{\url{hep-ph/0011074}}.

\bibitem{Giunti:2001kj}
C. Giunti,
\newblock Mod. Phys. Lett. A16 (2001) 2363,
  \href{http://arxiv.org/abs/hep-ph/0104148}{\url{hep-ph/0104148}}.

\bibitem{Nussinov:1976uw}
S. Nussinov,
\newblock Phys. Lett. B63 (1976) 201.

\bibitem{Kiers:1996zj}
K. Kiers, S. Nussinov and N. Weiss,
\newblock Phys. Rev. D53 (1996) 537,
  \href{http://arxiv.org/abs/hep-ph/9506271}{\url{hep-ph/9506271}}.

\bibitem{hep-ph/9305276}
C. Giunti et~al.,
\newblock Phys. Rev. D48 (1993) 4310,
  \href{http://arxiv.org/abs/hep-ph/9305276}{\url{hep-ph/9305276}}.

\bibitem{hep-ph/9709494}
C. Giunti, C.W. Kim and U.W. Lee,
\newblock Phys. Lett. B421 (1998) 237,
  \href{http://arxiv.org/abs/hep-ph/9709494}{\url{hep-ph/9709494}}.

\bibitem{hep-ph/9909332}
C.Y. Cardall,
\newblock Phys. Rev. D61 (2000) 073006,
  \href{http://arxiv.org/abs/hep-ph/9909332}{\url{hep-ph/9909332}}.

\bibitem{Beuthe:2002ej}
M. Beuthe,
\newblock Phys. Rev. D66 (2002) 013003,
  \href{http://arxiv.org/abs/hep-ph/0202068}{\url{hep-ph/0202068}}.

\bibitem{hep-ph/0312256}
C. Giunti,
\newblock (2003),
  \href{http://arxiv.org/abs/hep-ph/0312256}{\url{hep-ph/0312256}}.

\bibitem{Blasone:1995zc}
M. Blasone and G. Vitiello,
\newblock Ann. Phys. 244 (1995) 283,
  \href{http://arxiv.org/abs/hep-ph/9501263}{\url{hep-ph/9501263}}.

\bibitem{Blasone:1998hf}
M. Blasone, P.A. Henning and G. Vitiello,
\newblock Phys. Lett. B451 (1999) 140,
  \href{http://arxiv.org/abs/hep-th/9803157}{\url{hep-th/9803157}}.

\bibitem{Fujii:1998xa}
K. Fujii, C. Habe and T. Yabuki,
\newblock Phys. Rev. D59 (1999) 113003,
  \href{http://arxiv.org/abs/hep-ph/9807266}{\url{hep-ph/9807266}}.

\end{thebibliography}
\end{document}